\documentclass[11pt]{cernrep}
\usepackage{here}                                                                        \usepackage{graphicx}
\usepackage{booktabs}
\usepackage{cite}

\newcommand{\neut}{\ensuremath{\tilde{\chi}^0}}
\newcommand{\GeV}{\ensuremath{~\mathrm{GeV}}}
\newcommand{\MeV}{\ensuremath{~\mathrm{MeV}}}
\newcommand{\q}{\ensuremath{\mathrm{q}}}
\newcommand{\qbar}{\ensuremath{\bar{\mathrm{q}}}}
\renewcommand{\t}{\ensuremath{\mathrm{t}}}
\newcommand{\tbar}{\ensuremath{\bar{\mathrm{t}}}}
\renewcommand{\b}{\ensuremath{\mathrm{b}}}
\newcommand{\bbar}{\ensuremath{\bar{\mathrm{b}}}}
\newcommand{\fb}{\ensuremath{~{\mathrm{fb}}}}
\newcommand{\W}{\ensuremath{{\mathrm{W}}}}

\begin{document}

\title{MEASURING NEUTRINO MIXING ANGLES AT LHC}
\author{W.~Porod$^1$ and P.~Skands$^2$}
\institute{$^1$Institut f\"ur Theoretische Physik, University of Z\"urich,
           CH-8057 Z\"urich, Switzerland \\
$^2$Dept.~of Theoretical Physics,  Lund University, SE-223 62 Lund, Sweden
}
\maketitle
\begin{abstract}
We study an MSSM model with bilinear R-parity violation which is capable of
explaining neutrino data while leading to testable predictions for ratios of
LSP decay rates. Further, we estimate the precision with which such 
measurements could be carried out at the LHC.
\end{abstract}

\section{INTRODUCTION}

Recent neutrino experiments 
\cite{Fukuda:1998mi,Fukuda:2001nj,Ahmad:2001an,Eguchi:2002dm}
clearly show that neutrinos are massive particles and that they mix.
In supersymmetric models these findings can be explained by the
usual seesaw mechanism 
\cite{Gell-Mann:1980vs,Yanagida:1980xy,Mohapatra:1980ia}. 
However, supersymmetry allows for an
alternative which is intrinsically supersymmetric, namely the breaking
of R-parity.
The simplest way to realize this idea
 is to add bilinear terms to the superpotential $W$:
\begin{eqnarray}
W = W_{\rm MSSM} + \epsilon_i \hat L_i \hat H_u
\label{eq:model}
\end{eqnarray}
For consistency one has also to add the corresponding 
bilinear terms to soft SUSY breaking which induce small vacuum expectation
values (vevs) for the sneutrinos. These vevs in turn induce a mixing between
neutrinos and neutralinos, giving mass to one neutrino
at tree level. The second neutrino
mass is induced by loop effects 
(see \cite{Romao:1999up,Hirsch:2000ef,Diaz:2003as} 
and references therein). The same parameters
that induce neutrino masses and mixings are also responsible for the
decay of the lightest supersymmetric particle (LSP). This implies that there
are correlations between neutrino physics and LSP decays 
\cite{Mukhopadhyaya:1998xj,Porod:2000hv,Hirsch:2002ys,Hirsch:2003fe}.

In this note we investigate how well LHC can measure ratios of
LSP branching ratios that are correlated to 
 neutrino mixing angles in a scenario where
the lightest neutralino $\neut_1$ is the LSP.
In particular
we focus on the  semi-leptonic
final states $l_i \q' \qbar$ ($l_i=e,\mu,\tau$).
 There are several more
examples which are discussed in \cite{Porod:2000hv}.
In the model specified by Eq.~(\ref{eq:model})
the atmospheric mixing angle at tree level is given by
\begin{eqnarray}
 \tan \theta_{\rm atm} &=& \frac{\Lambda_2}{\Lambda_3} \\
\Lambda_i = \epsilon_i v_d + \mu v_i
\end{eqnarray}
where $v_i$ are the sneutrino vevs and $v_d$ is the vev of $H^0_d$. 
It turns out that the dominant part of the 
$\neut_1$-$\W$-$l_i$ coupling $O^L_i$ is given by
\begin{eqnarray}
 O^L_i = \Lambda_i f(M_1,M_2,\mu,\tan \beta, v_d, v_u)
\end{eqnarray}
where the exact form of $f$ can be found in Eq.~(20) 
of ref.~\cite{Porod:2000hv}. The important point is that $f$ only depends on
MSSM parameters but not on the R-parity violating parameters. Putting 
everything together one finds:
\begin{eqnarray}
 \tan^2 \theta_{\rm atm} \simeq \left| \frac{\Lambda_2}{\Lambda_3} \right|^2
   \simeq \frac{BR(\neut_1 \to \mu^\pm \W^\mp)}
               {BR(\neut_1 \to \tau^\pm \W^\mp)}
   \simeq  
 \frac{BR(\neut_1 \to \mu^\pm \qbar \q')}
      {BR(\neut_1 \to \tau^\pm \qbar \q' )},
\label{eq:corr}
\end{eqnarray}
where the last equality is only approximate due to possible (small)
contributions from three body decays of intermediate sleptons and squarks.
The restriction to the hadronic final states of the $\W$ is necessary
for the identification of the lepton flavour. Note that
Eq.~(\ref{eq:corr}) is a prediction of the bilinear model independent
of the R-parity conserving parameters.

\section{NUMERICAL RESULTS}

We take the SPS1a mSUGRA benchmark point \cite{Allanach:2002nj} as a
specific example, characterized by $m_0=100\GeV$, $m_\frac12=250\GeV$,
$A_0=-100\GeV$, $\tan\beta=10$, and $\mathrm{sign}(\mu)=1$\footnote{Strictly
  speaking, the SPS points should be defined by their low-energy parameters as
calculated with ISAJET 7.58.}. The
low--energy parameters were derived using \textsc{SPheno} 2.2
\cite{Porod:2003um} and passed to
\textsc{Pythia} 6.3 \cite{Sjostrand:2003wg} using the recently defined SUSY
Les Houches Accord \cite{Skands:2003cj}. The 
R-parity violating parameters (in \MeV) at the low scale are given by:
$\epsilon_1=43$, $\epsilon_2=100$, $\epsilon_3=10$, $v_1=-2.9$, $v_2=-6.7$
and $v_3=-0.5$. For the neutrino sector
we find $\Delta m^2_{\rm atm} = 3.8 \cdot 10^{-3}$~eV$^2$, 
$\tan^2 \theta_{\rm atm}= 0.91$, 
$\Delta m^2_{\rm sol} = 2.9 \cdot 10^{-5}$~eV$^2$,
$\tan^2 \theta_{\rm sol}= 0.31$. Moreover, we find that the following
neutralino branching ratios are larger than 1\%:
\begin{center}
\begin{tabular}{lll}
BR$(\W^\pm \mu^\mp) = 2.2\%$, & BR$(\W^\pm \tau^\mp) = 3.2\%$, &
 \\  BR$(\qbar \q' \mu^\mp) = 1.5\%$, & BR$(\qbar \q' \tau^\mp) = 2.1\%$, &
 BR$(\b \bbar \nu_i) = 15.6\%$, \\
BR$(e^\pm \tau^\mp \nu_i) = 5.9\%$, &BR$(\mu^\pm \tau^\mp \nu_i) = 30.3\%$, 
 &BR$(\tau^+ \tau^- \nu_i) = 37.3\%$, \\
\end{tabular}
\end{center}
where we have summed over the neutrino final states as well as over
the first two generations of quarks. Moreover, there are 0.2\% of neutralinos
decaying invisibly into three neutrinos. In the case that such events 
can be identified they can be used to distinguish this model from a model
with trilinear R-parity violating couplings because in the latter case they
are absent.

We now turn to the question to what extent
the ratio, Eq.~(\ref{eq:corr}), could be measurable at an LHC experiment. The
intention here is merely to illustrate the phenomenology and to give a rough idea of
the possibilities. For simplicity, we employ a number of shortcuts;
e.g.~detector energy resolution effects are ignored and events are only
generated at the parton level. Thus, we label a final-state quark or gluon
which has $p_\perp>15\GeV$ and which lies within the fiducial volume of the
calorimeter, $|\eta|<4.9$, simply as `a jet'. Charged leptons 
are required to lie within the inner detector coverage,
$|\eta|<2.5$, and to have $p_\perp>5\GeV$ (electrons), $p_\perp>6\GeV$
(muons), or $p_\perp>20\GeV$ (taus). 
The assumed efficiencies for such leptons are \cite{unknown:1999fq} 
75\% for electrons, 95\% for
muons, and 85\% for taus decaying in the 3--prong modes (we do not use the
1--prong decays), independent of $p_\perp$. 

For SPS1a, the total SUSY cross section is 
$\sigma_{\mathrm{SUSY}}\sim 41~\mathrm{pb}$. This consists mainly of gluino and squark
pair production followed by subsequent cascades down to the LSP, the
$\neut_1$. With an integrated luminosity of $100\fb^{-1}$, approximately 8 million $\neut_1$ decays should thus 
have occurred in the detector. 

An important feature of the scenario considered here is that the  
$\neut_1$ width is sufficiently small to result in a potentially observable
displaced vertex. By comparing the decay length, $c\tau =0.5$~mm, 
with an estimated vertex resolution of about 20 microns in the transverse
plane and 0.5 mm along the beam axis, it is apparent that the two neutralino
decay  
vertices should exhibit observable displacements in a fair
fraction of events. Specifically, we require that both neutralino
decays should occur outside an ellipsoid defined by 5 times the
resolution. For at least one of the vertices (the `signal' vertex), 
all three decay products ($\mu\q\qbar'$ or $\tau\q\qbar'$) must be
reconstructed,  while we only require one
reconstructed decay product (jet in the inner detector or lepton in the inner
detector whose track does not intersect the 5$\sigma$ vertex resolution ellipsoid)
for the second vertex (the `tag' vertex). 

Naturally, since the decay occurs within the detector, the standard SUSY
missing $E_\perp$ triggers are ineffective. Avoiding a discussion of detailed
trigger menus (cf.~\cite{tridas}), we have approached the issue by 
requiring that each event contains either four jets, each with $p_\perp >
100\GeV$, or two jets with $p_\perp>100\GeV$ together with a 
lepton (here meaning muon or electron) with $p_\perp>20\GeV$, or one jet with
$p_\perp>100\GeV$ together with two leptons with $p_\perp>20\GeV$. 
Further,
since the Standard Model background will presumably be dominated by $\t\tbar$
events, we impose an additional parton--level $\b$ jet veto. 

To estimate the efficiency with which decays into each channel can be
reconstructed, a sample of 7.9 million SUSY events were generated with
\textsc{Pythia}, and the above trigger and reconstruction cuts were
imposed. To be conservative, we only include the resonant decay channels,
where the quark pair at the signal vertex has the invariant mass of the $\W$.
The number of generated decays into each channel, the fractions
remaining after cuts, and the expected total number of 
reconstructed events scaled to an integrated luminosity of
$100\fb^{-1}$ are given in table \ref{tab:reconstruction}.
\begin{table}[tp]
\begin{center}
\begin{tabular}{lrr|r}
mode & $N_{gen}$ & $\epsilon_{\mathrm{rec}}$ & $N_{\mathrm{rec}}(100\fb^{-1})$\\ 
\toprule
$\neut_1\to \mu \W \to \mu\q\qbar'$ & 235000 & 0.10 & 12500  \\
$\neut_1\to \tau \W\to \tau_{\mathrm{3-prong}}\q\qbar'$ & 51600 & 0.054 & 1400
\\ \bottomrule 
\end{tabular}
\caption{Statistical sample, estimated reconstruction efficiencies, and expected event numbers. 
\label{tab:reconstruction}}
\end{center}
\end{table}
The comparatively small efficiencies owe mainly to the
requirement that \emph{both} neutralino decays should pass the 5$\sigma$
vertex resolution cut. Nonetheless, using these numbers as a first estimate, 
the expected statistical accuracy of the ratio, $R={BR(\neut_1 \to 
  \mu^\pm \W^\mp)} / {BR(\neut_1 \to \tau^\pm \W^\mp)}$, appearing in
  Eq.~(\ref{eq:corr}) becomes $\frac{\sigma(R)}{R} \simeq 0.028$.

\section{CONCLUSIONS}

We have studied neutralino decays in a model where bilinear R-parity violating
terms are added to the usual MSSM Lagrangian. This model can successfully
explain neutrino data and leads at the same time to {\it predictions}
for ratios of the LSP decay branching ratios. In particular we have considered
a scenario where the lightest neutralino is the LSP. In this case the  ratio 
${BR(\neut_1 \to \mu^\pm \W^\mp)}/{BR(\neut_1 \to \tau^\pm \W^\mp)}$ is
directly related to the  atmospheric neutrino mixing angle.  
Provided R-parity violating
 SUSY is discovered, the measurement of this ratio at colliders
would thus constitute an important test of the hypothesis of
a supersymmetric origin of neutrino masses. 

We have investigated the possibility of performing this measurement at
a `generic' LHC experiment, using \textsc{Pythia} to generate LHC SUSY
events at the parton level and imposing semi-realistic acceptance and
reconstruction cuts. Within this simplified framework, we find that
the LHC should be sensitive to a possible connection between 
 R-parity violating LSP decays  and
the atmospheric mixing angle, at least for scenarios with a fairly
light sparticle spectrum and where the neutralino decay length is
sufficiently large to give observable displaced vertices. Obviously,
the numbers presented here represent crude estimates and should not be
taken too literally. A more refined experimental analysis would be
necessary for more definitive conclusions to be drawn.

\vskip1cm
\noindent

\section*{ACKNOWLEDGEMENTS}

We thank A.~De Roeck and T.~Sj\"ostrand for useful discussions.
We are also grateful to the NorduGRID project for use of computing resources.
W.P.~is supported by the `Erwin Schr\"odinger fellowship No.~J2272'
of the `Fonds zur F\"orderung der wissenschaft\-lichen Forschung' of
Austria and partly by the Swiss `Nationalfonds'.

\bibliography{LH_BRPV_rev}

\end{document}